# ENTROPY PRODUCTION IN A GENERALIZED BREATHING PARABOLA MODEL: EXACT PATH INTEGRAL CALCULATIONS

Neha Tyagi and Binny J. Cherayil,[1]

Dept. of Inorganic and Physical Chemistry,

Indian Institute of Science,

Bangalore-560012, India

**Abstract**

Models of particle dynamics based on Brownian motion and its variants are a rich source of insights into the stochastic behaviour of complex condensed phase systems. In this paper we use one such variant – a breathing parabola with an additive time-dependent term $b(t)$ – as a non-trivial and previously unexplored model system for the verification of the integral fluctuation theorem (IFT). We demonstrate the IFT's applicability to this system within the framework of an exact path integral calculation. As a by-product of the calculation, we also show that in the limit $b(t) = 0$, where the model is representative of the solution dynamics of a colloid trapped in a harmonic potential with a time-dependent spring constant $a(t)$, the mean of the total entropy production $\Delta S_{tot}$ can be obtained in closed form as a function of $a(t)$. This result is expected to be relevant to the study of colloidal heat engines and other cyclically operating molecular machines. While $\Delta S_{tot}$ conforms to the IFT (and therefore assumes both positive and negative values), its mean is shown to increase monotonically with time, as required by the second law of thermodynamics.

[1] Corresponding author. Electronic mail: cherayil@iisc.ac.in

I. INTRODUCTION

The burgeoning field of stochastic thermodynamics – increasingly important in the study of small systems – often relies for its predictive and explanatory power on the theoretical insights provided by elementary models of many-body dynamics [1-4]. Few such models have likely contributed as much to the field as Brownian walks and their numerous variants, which are notable both for their wide applicability and their tractable mathematical properties. They have been used to study a host of different phenomena, and now serve as a standard paradigm of random behaviour [5].

The basic one-dimensional continuum version of the random walk model can be written

$$\frac{dx(t)}{dt} = \theta(t) \tag{1}$$

where $x(t)$ is some generic time-dependent variable (position, share price, chemical concentration, etc.) that changes its value according to the dynamics of a second variable $\theta(t)$, which fluctuates at random according to some prescribed rule. If $\theta(t)$ is white noise, for instance, $x(t)$ evolves as a simple random walk, and Eq. (1) can then describe the trajectory of a colloid in water [6,7] (along one Cartesian axis), or the rise and fall of stock market indices [8], or any of a number of other events that vary irregularly in time. The addition of other terms to the equation, such as a term linear in $x(t)$, gives it even greater versatility, allowing it, for example, to now describe a *trapped* colloid in an aqueous medium [9-12].

Still more general forms of Eq. (1) are possible, such as the equation below,

$$\frac{dx(t)}{dt} = b(t) - a(t)x(t) + \xi(t) \qquad (2)$$

where $a(t)$ and $b(t)$ are arbitrary functions of time. This equation can also serve as a model of various dynamical phenomena. For instance, if $b(t) = 0$, it can describe the motion of a colloid in an optical trap whose strength *changes* with time, leading to what is referred to as a "breathing parabola" model [13-17]. The case $b(t) \neq 0$ then describes such a colloid in the presence of an external time-dependent force, which is the subject of this paper.

The above applications of Eq. (2) describe situations that typically evolve under far-from-equilibrium conditions. Thermodynamic behaviour in such conditions is now known to be governed by a variety of fluctuation theorems, which are statements about the distributions of fluctuating quantities like heat, work and entropy [18-21]. These theorems are of considerable scope and generality, and are widely applicable. Several such theorems have, in fact, been shown to be true of systems governed by special cases of the dynamics defined by Eq. (2), including the following: (i) $b(t) = 0$ and $a(t)$ arbitrary [22] (a generalization of the case $b(t) = 0$ and $a(t)$ either a step function or a sinusoidally varying function [23]), and (ii) $b(t)$ arbitrary and $a(t)$ a constant [24,25], (a generalization of the case $a(t)$ a constant and $b(t) \propto t$ [22,26]). But to the best of our knowledge, the case in which both $a(t)$ and $b(t)$ are arbitrary has still not been explored, and we believe it would be of considerable interest (both pedagogically and from the point of abstract theory) to test – *analytically* – the validity of one or other of the fluctuation theorems for this case. In the present paper, with this end in view, we show how a path integral formalism can be used to prove that Eq. (2) satisfies the so-called integral fluctuation theorem, a relation connected to the total entropy produced by a system and its surroundings in the presence of external driving. In doing so, we illustrate the application of modern concepts in stochastic thermodynamics to a model that is highly non-trivial but that is nevertheless amenable to *exact* analytical treatment.

The organization of the paper is as follows. Sections II A, B and C discuss how the energy, work, heat and entropy of our model oscillator system are related to its stochastic trajectories. Section II C provides an exact expression – obtained from a path integral formalism – for the conditional probability density distribution (i.e., propagator) for these trajectories. In Sec. II D, this propagator is used to demonstrate that the exponential average of the total entropy production of the system and its surroundings satisfies the integral fluctuation theorem. Section III discusses the calculation of the mean of the total entropy production for the case $b(t) = 0$ (i.e., for the breathing parabola model), and for a special choice of the protocol that drives the system out of equilibrium. Section IV is a summary of the main findings of our study.

## II. STATISTICAL THERMODYNAMICS OF THE OSCILLATOR MODEL

### A. BACKGROUND

For the present purposes, we shall interpret Eq. (2) – the defining equation of our model – as the equation of motion of a point particle located at $x$ at time $t$ that moves in one dimension under the influence of forces arising from (i) a time-dependent potential $U$ given by

$$U = \frac{1}{2} a(t) x^2(t) - b(t) x(t), \qquad (3)$$

where *a* and *b* are now understood to refer, respectively, to the ratios $k/\zeta$ and $F/\zeta$, with *k* a spring constant, *F* a force and $\zeta$ a friction coefficient, and (ii) a heat bath at constant temperature *T* whose fluctuations are modelled by $\xi(t)$, which is treated as a white noise variable with the correlations $\langle \xi(t) \rangle = 0$ and $\langle \xi(t)\xi(t') \rangle = 2D\delta(t-t')$, where *D*, by virtue of the fluctuation-dissipation theorem, is given by $D = k_B T/\zeta \equiv 1/\beta\zeta$, $k_B$ being Boltzmann's constant. In this interpretation, the particle can be regarded as a single-molecule system interacting through a porous boundary with a thermal reservoir, and by so doing, exchanging energy, work and heat with it. These thermodynamic quantities are fluctuating variables themselves, and depend sensitively on initial conditions and particle trajectories. So in different repetitions of a protocol that transfers the system between *nominally* identical initial and final states, they typically do not assume the same values. But averages of certain trajectory-dependent variables over the *spread* of these values often satisfy one or more well-defined mathematical relations. These relations are the fluctuation theorems alluded to in the Introduction [18-21], of which the integral fluctuation theorem (IFT) is especially notable, being in some sense the extension to the molecular realm of the second law of thermodynamics. The IFT takes the form

$$\langle e^{-\Delta S_{tot}/k_B} \rangle = 1 \qquad (4)$$

where $\Delta S_{tot}$ is the change in entropy of system *and* surroundings in the time interval *t*, and the angular brackets denote an average over the distributions both of the particle's initial position and its stochastic trajectories.

**B. HEAT AND WORK**

To establish whether the model defined by Eq. (2) satisfies Eq. (4) for any general functions $a(t)$ and $b(t)$, it is necessary to define appropriate stochastic analogues of the thermodynamic quantities that characterize macroscopic systems during changes of state. One of these quantities is the work, $W$, which is generally identified with the change in internal energy $\Delta U$ induced by a change in a time-dependent control parameter [18-21]. For the model of Eq. (2), with $U$ given by Eq. (3), the work done, $W(t)$, in an interval of time $t$, is

$$W(t) = \int_0^t dt' \frac{\partial U}{\partial t'} = \zeta \int_0^t dt' \left( \frac{1}{2} \dot{a}(t') x^2(t') - \dot{b}(t') x(t') \right) \tag{5}$$

where the dots on $a$ and $b$ denote a derivative with respect to time. If the value of $x$ at the end-points 0 and $t$ of the given time interval are denoted $x_0$ and $x_f$, respectively, with $b(0)$ chosen to be identically 0, $\Delta U$ itself is given by

$$\Delta U = \zeta \left[ -b(t) x_f + \frac{1}{2} (a(t) x_f^2 - a(0) x_0^2) \right] \tag{6}$$

By the first law of thermodynamics, the heat $Q(t)$ dissipated into the medium during the change of state is $Q(t) = W(t) - \Delta U$, and so, from Eqs. (5) and (6)

$$Q(t) = \zeta \int_0^t dt' \left( \frac{1}{2} \dot{a}(t') x^2(t') - \dot{b}(t') x(t') \right) + \zeta \left[ b(t) x_f - \frac{1}{2} (a(t) x_f^2 - a(0) x_0^2) \right]. \tag{7}$$

Assuming that this heat is manifested as a change in entropy of the medium, $\Delta S_m$, it follows that $\Delta S_m = Q(t)/T$.

## C. SYSTEM ENTROPY

To calculate the change in entropy of the *system*, $\Delta S$, (which we need in the calculation of $\Delta S_{tot}$, which is defined as $\Delta S_m + \Delta S$) we adopt the prescription introduced by Seifert [27], which relates $\Delta S$ to the stochastic trajectory of the particle through the expression $\Delta S = -k_B \ln P(x_f, t)/P_{eq}(x_0)$, where $P(x_f, t)$ is the probability density that at time *t*, the particle is at the point $x_f$, while $P_{eq}(x_0)$ is the probability density that at time 0, it is at $x_0$. $P(x_f, t)$ in turn can be written quite generally as $P(x_f, t) = \int_{-\infty}^{\infty} dx_0 P(x_f, t | x_0) P_{eq}(x_0)$, where $P(x_f, t | x_0)$ is the conditional probability density that the particle is at $x_f$ at time *t* given that it was at $x_0$ at time $t = 0$. The derivation of an expression for this conditional probability, and from there an expression for $P(x_f, t)$ and later $\Delta S$, is a critical element of the present calculations, and it is the subject we turn to next.

Given that $\xi(t)$ in Eq. (2) is a Gaussian random variable with zero mean, the probability $P[\xi]$ that it follows a particular stochastic trajectory in the time interval *t* is proportional to the functional $\exp\left[-(1/4D)\int_0^t dt' \xi^2(t')\right]$. This means that the corresponding trajectory of the variable *x* occurs with a probability $P[x]$ that is given by [28]

$$P[x] = \mathcal{N} J[x] \exp\left[-\frac{1}{4D}\int_0^t dt'(\dot{x}^2 + b^2 + a^2 x^2 - 2b\dot{x} + 2a\dot{x}x - 2abx)\right] \quad (8)$$

where $\mathcal{N}$ is a normalization constant (to be fixed later) and $J[x]$ is the Jacobian for the transformation from $\xi$ to *x* variables, which can be shown, using the discretization approach discussed in Ref. [28] to produce $J = e^{(1/2)\int_0^t dt' a(t')}$. After further simplification, Eq. (8) can be expressed as

$$P[x] = \mathcal{N} e^{-\mathcal{A}} \exp\left[\frac{1}{2}\int_0^t dt' a(t') - \frac{1}{4D}\int_0^t dt' b^2(t') + \frac{1}{4D}\left(2b(t)x_f - a(t)x_f^2 + a(0)x_0^2\right)\right] \quad (9a)$$

where the "action" $\mathcal{A}$ is defined as $\mathcal{A} = \int_0^t dt' \mathcal{L}(\dot{x}, x)$, the "Lagrangian" $\mathcal{L}$ being given by

$$\mathcal{L} = \frac{1}{4D}\left[\dot{x}^2(t') + p(t')x^2(t') + 2q(t')x(t')\right] \quad (9b)$$

with $p(t') \equiv a^2(t') - \dot{a}(t')$ and $q(t') \equiv -a(t')b(t') + \dot{b}(t')$. The function $P(x_f, t | x_0)$ can now be written as

$$P(x_f, t | x_0) = \mathcal{N}\pi(x_f, t | x_0)\exp\left[-\frac{1}{4D}\left(a(t)x_f^2 - a(0)x_0^2 - 2b(t)x_f - 2D\int_0^t dt' a(t')\right.\right.$$

$$\left.\left. + \int_0^t dt' b^2(t')\right)\right] \quad (10)$$

where the propagator $\pi(x_f, t | x_0)$ is the path integral

$$\pi(x_f, t | x_0) = \int_{x(0)=x_0}^{x(t)=x_f} \mathcal{D}[x]\exp\left[-\int_0^t dt' \mathcal{L}(\dot{x}, x)\right] \quad (11)$$

By finding the path $\bar{x}(t)$ that minimizes $\mathcal{A}$ and then evaluating $\mathcal{A}$ along it, this propagator, by virtue of being a quadratic functional of $x$, can be determined exactly as $\pi(x_f, t | x_0) \propto \mathcal{M}(t) e^{-\bar{\mathcal{A}}}$, where $\bar{\mathcal{A}}$ is the value of $\mathcal{A}$ along $\bar{x}(t)$, and $\mathcal{M}(t)$ is a path-independent coefficient that depends solely on $t$ and that can be determined from the deWitt-Morette formula [28,29] $\mathcal{M}(t) = \sqrt{|\partial^2 \overline{\mathcal{A}} / \partial x_f \partial x_0|}$. The proportionality constant in the above expression for $\pi(x_f, t | x_0)$ is combined with $\mathcal{N}$, and the product, $\mathcal{N}'$, is then found from the requirement that $\int_{-\infty}^{\infty} dx_f P(x_f, t | x_0)$ be unity.

For $\mathcal{A}$ to be a minimum, $\bar{x}(t)$ must satisfy the following Euler-Lagrange equation: $(d/dt)\partial \mathcal{L}/\partial \dot{x} = \partial \mathcal{L}/\partial x$, i.e., it must satisfy

$$\ddot{\bar{x}}(t) - p(t)\bar{x}(t) = q(t) \tag{12}$$

The solution of this equation is [30]

$$\bar{x}(\tau) = \int_0^t dt' G(t', \tau) q(t') + x_f \frac{d}{dt'} G(t', \tau) \bigg|_{t'=t} - x_0 \frac{d}{dt'} G(t', \tau) \bigg|_{t'=0} \tag{13}$$

where $G$ is a Green's function that is obtained from the equation

$$\left(\frac{d^2}{dt^2} - p(t)\right) G(t, t') = \delta(t - t') \tag{14}$$

under the boundary conditions $G(0, t') = G(t, t') = 0$. From the solution to the homogeneous equation $(d^2/dt^2 - p(t))G(t, t') = 0$, and from the continuity and boundary conditions on $G$, one can show that $G$ is given by

$$G(t', t'') = \begin{cases} -\psi(t')\phi(t'') + \dfrac{\phi(t)}{\psi(t)}\psi(t')\psi(t''), & t' < t'' \\ -\phi(t')\psi(t'') + \dfrac{\phi(t)}{\psi(t)}\psi(t'')\psi(t'), & t' > t'' \end{cases} \tag{15}$$

where $\phi(t) = \exp\left(-\int_0^t dt' a(t')\right)$ and $\psi(t) = \phi(t)\int_0^t dt' \phi(t')^{-2}$.

Using integration by parts, the minimized action, $\bar{\mathcal{A}} = \mathcal{A}(\bar{x}) = (1/4D)\int_0^t dt'[\dot{\bar{x}}^2(t') + p(t')\bar{x}^2(t') + 2q(t')\bar{x}(t')]$, can now be written as

$$\bar{\mathcal{A}} = \frac{1}{4D}[\dot{\bar{x}}(t)x_f - \dot{\bar{x}}(0)x_0 + \int_0^t dt'\bar{x}(t')q(t')] \tag{16}$$

After introducing the expression for $\bar{x}(\tau)$ from Eq. (13) into the above equation, substituting the result into the expression for the propagator $\pi(x_f, t | x_0)$, substituting that result into the expression for $P(x_f, t | x_0)$, and then determining the value of the coefficient $\mathcal{N}'$ by carrying out the lengthy but relatively straightforward integration over $x_f$ in the normalization condition $\int_{-\infty}^{\infty} dx_f P(x_f, t | x_0) = 1$, one can finally show (after finding that $\mathcal{N}' = 1/\sqrt{2\pi}$) that

$$P(x_f, t | x_0) = \frac{e^{\Lambda(t)}}{\sqrt{4\pi D\psi(t)}} \exp\left[-\left\{\frac{x_f^2}{4D\psi(t)\phi(t)} + \frac{x_0^2\phi(t)}{4D\psi(t)} - \frac{x_f x_0}{2D\psi(t)} - x_f\Xi_1(t) - x_0\Xi_2(t)\right\}\right]$$

(17a)

where

$$\Lambda(t) = \frac{1}{2}\int_0^t dt' a(t') - \frac{1}{4D}\int_0^t dt' b^2(t') - \frac{1}{4D}\int_0^t dt'\int_0^t dt'' q(t')G(t'',t')q(t'') \tag{17b}$$

$$\Xi_1(t) = \frac{1}{2D\psi(t)}\left[b(t)\psi(t) - \int_0^t dt'\psi(t')q(t')\right] \tag{17c}$$

and

$$\Xi_2(t) = \frac{1}{2D\psi(t)}\int_0^t dt' q(t')[\phi(t)\psi(t') - \phi(t')\psi(t)] \tag{17d}$$

To proceed with the calculation of the distribution $P(x_f, t)$, we recall that in the initial state, at time $t = 0$, the system was assumed to be in equilibrium with the thermal reservoir at temperature $T$. The positions of the particle in this state are therefore governed by the Boltzmann distribution

$$P_{eq}(x_0) = \sqrt{\frac{a(0)}{2\pi D}} e^{-a(0)x_0^2/2D} \tag{18}$$

Multiplication of this expression by Eq. (17a), followed by integration of the product over $x_0$ from $-\infty$ to $+\infty$ yields $P(x_f, t)$, from which, following the Seifert prescription, the entropy change of the system, $\Delta S$, can be shown to be given by

$$\frac{\Delta S}{k_B} = -\Lambda(t) - \frac{D\psi(t)\Xi_2^2(t)}{\phi(t) + 2a(0)\psi(t)} + \frac{1}{2}\ln(\phi(t) + 2a(0)\psi(t)) - \frac{a(0)x_0^2}{2D} - x_f \Xi_1(t) +$$

$$+ \frac{1}{2D(\phi(t) + 2a(0)\psi(t))}\left[\frac{a(0)x_f^2}{\phi(t)} - \frac{x_f}{\psi(t)}\int_0^t dt' q(t')[\phi(t)\psi(t') - \phi(t')\psi(t)]\right] \tag{19}$$

## D. VALIDATING THE IFT

The total entropy produced by the system and surroundings, $\Delta S_{tot}$, can now be obtained from Eqs. (7) and (19). The exponential average, $\langle e^{-\Delta S_{tot}/k_B} \rangle$, is then found to be expressible as

$$\langle e^{-\Delta S_{tot}/k_B} \rangle = \frac{\sqrt{a(0)}}{\sqrt{2\pi D(\phi(t) + 2a(0)\psi(t))}} \exp\left[2\Lambda(t) + \frac{1}{4D}\int_0^t dt' \int_0^t dt'' q(t') G(t'', t') q(t'') + \right.$$

$$+ \frac{D\psi(t)\Xi_2^2(t)}{\phi(t)+2a(0)\psi(t)}\Bigg]I_1 \qquad (20a)$$

where $I_1$ is the integral

$$I_1 = \int_{-\infty}^{\infty} dx_f \int_{-\infty}^{\infty} dx_0 \exp\Bigg[-\frac{a(0)x_0^2}{4D} - \frac{x_f^2}{4D}\left(\frac{2a(0)}{\phi(t)(\phi(t)+2a(0)\psi(t))} - a(t)\right) + x_f$$

$$\left(\frac{\Xi_2(t)}{\phi(t)+2a(0)\psi(t)} - \frac{1}{2D\psi(t)}\int_0^t dt' q(t')\psi(t')\right)\Bigg]\pi_1(x_f,t\mid x_0) \qquad (20b)$$

with $\pi_1(x_f,t\mid x_0) \equiv \int_{x(0)=x_0}^{x(t)=x_f}\mathcal{D}[x]\exp[-\int_0^t dt'\mathcal{L}_1(\dot{x},x)] \equiv \int_{x(0)=x_0}^{x(t)=x_f}\mathcal{D}[x]e^{-\mathcal{A}_1}$, where $\mathcal{A}_1$ is the action, and $\mathcal{L}_1 \equiv [\dot{x}^2(t') + p_1(t')x^2(t') + 2q_1(t')x(t')]/4D$ is the Lagrangian, with $p_1(t') \equiv a^2(t') + \dot{a}(t')$ and $q_1(t') \equiv -a(t')b(t') - \dot{b}(t')$. (Notice that $\mathcal{L}_1$ differs from the earlier Lagrangian $\mathcal{L}$ [see Eq. (9b)] only by factors of $-1$.) The propagator $\pi_1$ is found using the same variational procedure described earlier, and assumes the form $\pi_1 = \mathcal{K}(t)e^{-\overline{\mathcal{A}}_1}/\sqrt{2\pi}$, where $\overline{\mathcal{A}}_1$ is value of the action evaluated along the classical trajectory, and $\mathcal{K}(t) = \sqrt{|\partial^2 \overline{\mathcal{A}}_1/\partial x_f \partial x_0|}$. The classical trajectory is found from the solution to the Euler-Lagrange equation $\ddot{\bar{x}}(t') - p_1(t')\bar{x}(t') = q_1(t')$. This solution can be shown to be (see Eq. (13))

$$\bar{x}(\tau) = \int_0^t dt' G_1(t',\tau)q_1(t') + x_f \frac{\Psi(\tau)}{\Psi(t)} - \frac{x_0}{\Psi(t)}(\Phi(t)\Psi(\tau) - \Psi(t)\Phi(\tau)) \qquad (21a)$$

where $\Phi(t) = \exp\left(+\int_0^t dt' a(t')\right)$, $\Psi(t) = \Phi(t)\int_0^t dt' \Phi(t')^{-2}$, and the new Green's function $G_1(t',t'')$ is given by

$$G_1(t',t'') = \begin{cases} -\Psi(t')\Phi(t'') + \dfrac{\Phi(t)}{\Psi(t)}\Psi(t')\Psi(t''), & t' < t'' \\ -\Phi(t')\Psi(t'') + \dfrac{\Phi(t)}{\Psi(t)}\Psi(t'')\Psi(t'), & t' > t'' \end{cases} \quad (21b)$$

The minimized action $\overline{\mathcal{A}}_1$ is found by using the above expression for $\overline{x}(\tau)$ in

$$\overline{\mathcal{A}}_1 = [\dot{\overline{x}}(t)x_f - \dot{\overline{x}}(0)x_0 + \int_0^t dt'\overline{x}(t')q_1(t')]/(4D),$$ (cf. Eq. (16)), which leads to

$$\overline{\mathcal{A}}_1 = \frac{1}{4D}\left[\frac{x_f^2}{\Psi(t)\Phi(t)}(1+a(t)\Psi(t)\Phi(t)) + \frac{x_0^2}{\Psi(t)}(\Phi(t)-a(0)\Psi(t)) - \frac{2x_0 x_f}{\Psi(t)} + \frac{2x_f}{\Psi(t)}\right.$$

$$\left.\int_0^t dt' q_1(t')\Psi(t') - \frac{2x_0}{\Psi(t)}\int_0^t dt' q_1(t')[\Phi(t)\Psi(t')-\Psi(t)\Phi(t')] + \int_0^t dt'\int_0^t dt'' q_1(t')G_1(t'',t')q_1(t'')\right]$$

(22)

From this result it follows that $\mathcal{K}(t) = 1/\sqrt{2D\Psi(t)}$.

The calculation of $\langle e^{-\Delta S_{tot}/k_B}\rangle$ is now completed by carrying out the Gaussian integrals over $x_f$ and $x_0$, collecting terms and simplifying the result. The calculations are not difficult, but they are involved, and some effort is needed to manipulate the terms that emerge from the integrations into forms that allow simplifying patterns among them to be recognized. An illustration of the kinds of manipulations that are used in the calculations is provided in Appendix A. What we eventually find is that $\langle e^{-\Delta S_{tot}/k_B}\rangle = 1$, verifying the IFT for this system.

## III. THE MEAN ENTROPY PRODUCTION

As mentioned in the Introduction, when the source term $b(t)$ in the generalized Ornstein-Uhlenbeck process [cf. Eq. (2)] is set to 0, the resulting equation describes the stochastic evolution of a so-called breathing parabola (also referred to in [23] as the "smoothly squeezed harmonic oscillator"), and can therefore serve as a model of the dynamics of a Brownian colloid in an optical trap of time-varying stiffness. A colloidal system of this kind has recently been used to construct a micron-sized Stirling engine [31,32], the adjustable stiffness of the trap providing the means to carry out the analogues of the isothermal expansion and compression steps in the macroscopic version of this engine.

The mean of the total entropy production, $\langle \Delta S_{tot} \rangle$, for the $b(t) = 0$ case has been calculated in Ref. [23] for the choice $a(t) = \sin^2(\pi t) + 1$. To complement this study, we present an expression for $\langle \Delta S_{tot} \rangle / k_B$ in the $b(t) = 0$ limit that is applicable to *any* choice of $a(t)$. Appendix B outlines the derivation of this expression, which, introducing the notation $\bar{\sigma}(t) \equiv \langle \Delta S_{tot} \rangle / k_B$, is given by

$$\bar{\sigma}(t) = \frac{1}{2a(0)} \int_0^t dt' \dot{a}(t') \left( \phi^2(t') + 2a(0)\phi(t')\psi(t') \right) + \frac{1}{2} \ln[\phi(t) + 2a(0)\psi(t)]$$

$$-\frac{1}{2}\left( \int_0^t dt' a(t') + 1 \right) + \frac{a(0)}{2a(t)(\phi^2(t) + 2a(0)\phi(t)\psi(t))} \quad (23)$$

For purposes of illustration, we now consider two forms of $a(t)$, one defined as

$$a(t) = \frac{\alpha}{1 + \mu t} \quad (24a)$$

and the other (which can be regarded as its "reverse") as

$$a_R(t) = \frac{\alpha}{1 + \mu(\tau - t)} \tag{24b}$$

where $\alpha$, $\mu$ and $\tau$ are arbitrary adjustable parameters. We next define a protocol where, starting from $t = 0$, a Brownian particle is allowed to evolve for a time $\tau$ according to Eq. (2), with $b(t)$ set to 0, and $a(t)$ fixed by Eq. (24a). During this protocol, the particle's dynamics takes place in a trapping potential of progressively *decreasing* stiffness. The mean total entropy, $\bar{\sigma}(t)$, that is generated during the particle's time evolution is found in closed form from Eq. (23) as a function of $\alpha$ and $\mu$. The final expression for $\bar{\sigma}(t)$ is somewhat lengthy and not especially instructive, so in the interests of brevity it is not reproduced here; instead we show the variation of $\bar{\sigma}(t)$ with $t$ graphically (see Fig. 1) after setting $\alpha$ to 1 and $\mu$ to 3 (these being entirely arbitrary values.)

The figure is divided into four adjacent panels, each separated by dashed lines, and it is the first panel that pertains to the protocol described above. The coloured lines here correspond to the time dependence of (i) the function $a(t)$ (blue) and (ii) the mean of the total entropy (red). It is evident from the figure that in general, after starting out at 0, $\bar{\sigma}(t)$ increases monotonically with $t$, as expected.

In a second protocol, the Brownian particle is assumed to start out once more from $t = 0$, but with the stiffness of its trapping potential now fixed by the function $a_R(t)$ of Eq. (24b), and $\tau$ set to the (arbitrary) value of 10 at which the first protocol was concluded. The particle is then allowed to evolve for a further period of $\tau$, so that at the end of this time $a(t) = \alpha$, the value it had at the start of the first protocol (the particle's evolution having taken place in a potential of progressively *increasing* stiffness.) During this second protocol, $\bar{\sigma}(t)$, as calculated from Eq. (23), starts out from 0 as before, and then increases monotonically with *t*. By adding the entropy generated at time $t = \tau$ in the first protocol (i.e., along the "forward" trajectory) to the entropy generated in the second protocol (along the "backward" trajectory), one can account for the net (or "running") entropy production over a period of time in which these two protocols are imagined to follow each other in succession. This running entropy production is the red curve in the second panel of Fig. 1, while the green curve in the same panel is the time variation of the function $a_R(t)$. These two protocols are then repeated, one after the other, so that physically the particle is subjected successively to a trap of decreasing stiffness and then one of increasing stiffness. The running entropy generated during the course of these changes is the red curve in panels 3 and 4, while the blue and green curves in these panels are, respectively, the functions $a(t)$ and $a_R(t)$. The variation of $\bar{\sigma}(t)$ with *t* in Fig. 1 mirrors the behaviour of this function in the work of Spinney and Ford [23] for the case $b(t) = 0$, $a(t) = \sin^2(\pi t) + 1$.

## IV. SUMMARY

In this paper we have demonstrated that the source-dependent breathing parabola model satisfies the integral fluctuation theorem for *arbitrary* functional forms of both the source term $b(t)$ and the stiffness parameter $a(t)$. This enlarges the set of models for which the exponential average of the total entropy production can be evaluated exactly [33-36], thereby not only providing a useful pedagogical illustration of how fluctuation theorems emerge, but also placing the physics of the model itself on firmer theoretical footing. These contributions to our understanding of the breathing parabola model assume significance in light of experimental and theoretical work that is being carried out on the thermodynamics of nanoscale heat engines [31,32,37,38], which frequently operate by delivering work and heat cyclically to colloid-based systems using optical traps of adjustable time-varying stiffness. We have also shown that during a protocol in which the stiffness of the potential confining a Brownian particle alternately decreases and increases (as in the operation of a colloid-based engine), the mean of the total entropy production increases monotonically, in accordance with the second law of thermodynamics.

**Acknowledgements** Part of this work was initiated by BJC in 2015 during a six-month stay as a Visiting Professor in the Department of Chemistry and Chemical Biology at Harvard University. Financial assistance for this visit was provided by Harvard University, and is gratefully acknowledged. The visit would not have been possible without the generous efforts of Eugene Shakhnovich and Xiao-Wei Zhuang of the Chemistry and Chemical Biology Department, to whom the author is especially grateful.

**APPENDIX A. REDUCTION OF CERTAIN INTEGRALS TO SIMPLER FORM**

As an example of the kinds of integrals that must be evaluated to determine the value of $\left\langle e^{-\Delta S_{tot}/k_B} \right\rangle$, consider $J_1 \equiv \int_0^t dt' \int_0^{t'} dt'' q(t') G(t'',t') q(t'')$, which comes from the second term in the argument of the exponential in Eq. (20a) after the integration range of the variable $t''$ is separated into two intervals, one from 0 to $t'$ and the other from $t'$ to $t$, a separation necessitated by the different functional forms of the Green's function in these intervals (though the two resulting integrals eventually yield the same result.)

The evaluation of $J_1$ proceeds by substituting into its definition the expression for $G$ from Eq. (15). This yields

$$J_1 = -\int_0^t dt' q(t')\phi(t') \int_0^{t'} dt'' q(t'')\psi(t'') + \frac{\phi(t)}{\psi(t)} \int_0^t dt' q(t')\psi(t') \int_0^{t'} dt'' q(t'')\psi(t'') \tag{A.1}$$

Recalling that $q(t) = -a(t)b(t) + \dot{b}(t)$, one then uses integration by parts to carry out the integration over $t''$, obtaining

$$\int_0^{t'} dt'' q(t'')\psi(t'') = b(t')\psi(t') - \int_0^{t'} dt'' b(t'')\phi(t'')^{-1}, \tag{A.2}$$

which is substituted back into Eq. (A.1). Integration by parts is used again in the resulting expression, this time in conjunction with the fact that the time derivatives of $\phi(t)$ and $\psi(t)$ are given by $\dot{\phi}(t) = -a(t)\phi(t)$ and $\dot{\psi}(t) = -a(t)\psi(t) + \phi(t)^{-1}$. In this way, one arrives at the following expressions for various integrals that occur in intermediate steps of the overall calculation of $J_1$:

$$\int_0^t dt' q(t')\phi(t')b(t')\psi(t') = \frac{1}{2}b^2(t)\phi(t)\psi(t) - \frac{1}{2}\int_0^t dt' b^2(t) \tag{A.3}$$

$$\int_0^t dt' q(t')\psi(t')b(t')\psi(t') = \frac{1}{2}b^2(t)\psi^2(t) - \int_0^t dt' b^2(t')\psi(t')\phi(t')^{-1} \tag{A.4}$$

$$\int_0^t dt' q(t')\phi(t') \int_0^{t'} dt'' b(t'') \phi(t'')^{-1} = b(t)\phi(t) \int_0^t dt' b(t')\phi(t')^{-1} - \int_0^t dt' b^2(t') \qquad (A.5)$$

and

$$\int_0^t dt' q(t')\psi(t') \int_0^{t'} dt'' b(t'') \phi(t'')^{-1} = b(t)\psi(t) \int_0^t dt' b(t')\phi(t')^{-1} - \frac{1}{2}\left(\int_0^t dt' b(t')\phi(t')^{-1}\right)^2$$

$$- \int_0^t dt' b^2(t')\psi(t')\phi(t')^{-1} \qquad (A.6)$$

After all these terms are collected together, the final expression for $J_1$ simplifies to

$$J_1 = -\frac{1}{2}\int_0^t dt' b^2(t') + \frac{\phi(t)}{2\psi(t)}\left(\int_0^t dt' b(t')\phi(t')^{-1}\right)^2 \qquad (A.7)$$

Other such integrals – listed below – are produced during the course of evaluating $\langle e^{-\Delta S_{tot}/k_B} \rangle$ via Eqs. (20a), (20b) and (22), and they are treated in much the same way. Omitting details of the calculations, we simply note that these integrals can be reduced to the following expressions:

$$J_2 \equiv \int_0^t dt' \int_0^{t'} dt'' q_1(t') G_1(t'',t') q_1(t'') = -\frac{1}{2}\int_0^t dt' b^2(t') + \frac{\Phi(t)}{2\Psi(t)}\left(\int_0^t dt' \frac{b(t')}{\Phi(t')}\right)^2 \qquad (A.8)$$

$$J_3 \equiv \int_0^t dt' \int_0^t dt'' q(t')q(t'')[\phi(t)\psi(t') - \phi(t')\psi(t)][\phi(t)\psi(t'') - \phi(t'')\psi(t)]$$

$$= \phi^2(t)\left(\int_0^t dt' \frac{b(t')}{\phi(t')}\right)^2 \qquad (A.9)$$

$$J_4 \equiv \int_0^t dt' \int_0^t dt'' q_1(t') q_1(t'') [\Phi(t)\Psi(t') - \Phi(t')\Psi(t)][\Phi(t)\Psi(t'') - \Phi(t'')\Psi(t)]$$

$$= \Phi^2(t) \left( \int_0^t dt' \frac{b(t')}{\Phi(t')} \right)^2 \tag{A.10}$$

$$J_5 \equiv \int_0^t dt' q(t')\psi(t') = b(t)\psi(t) - \int_0^t dt' \frac{b(t')}{\phi(t')} \tag{A.11}$$

$$J_6 \equiv \int_0^t dt' q_1(t')\Psi(t') = -b(t)\psi(t) + \int_0^t dt' \frac{b(t')}{\Phi(t')} \tag{A.12}$$

After all these contributions to $\langle e^{-\Delta S_{tot}/k_B} \rangle$ are collected together (cf. Eqs. (20a), (20b) and (22)), and the result simplified, it can be shown that $\langle e^{-\Delta S_{tot}/k_B} \rangle = 1$.

## APPENDIX B. CALCULATION OF THE MEAN OF THE TOTAL ENTROPY PRODUCTION

From Eqs. (7) and (19), we see that $\sigma(t) \equiv \Delta S_{tot}(b=0)/k_B$ is given by

$$\sigma(t) = \frac{1}{2D} \int_0^t dt' \dot{a}(t') x^2(t') + \Omega(t) \tag{B.1a}$$

where

$$\Omega(t) = \frac{1}{2} \ln[\phi(t) + 2a(0)\psi(t)] - \frac{1}{2} \int_0^t dt' a(t') - \frac{x_f^2}{2D} \left( a(t) - \frac{a(0)}{(\phi^2(t) + 2a(0)\phi(t)\psi(t))} \right)$$

(B.1b)

The above expression for $\sigma(t)$ agrees exactly with the expression for this quantity derived in Ref. [23] (cf. Eq. (1.169) in that reference.) The mean of $\sigma(t)$ is calculated by averaging (B.1) over both the realizations of the random variable $x(t)$ and over the distribution of $x_f$. Averaging over $x_f$ ensures that the system is in equilibrium in its final state [23], and it is carried out using the following form of the probability density function $P(x_f, t)$:

$$P(x_f, t) = \sqrt{\frac{a(t)}{2\pi D}} \exp\left(-a(t) x_f^2 / 2D\right) \tag{B.2}$$

Thus the mean of $\sigma(t)$ assumes the form

$$\bar{\sigma}(t) = \frac{1}{2D}\int_0^t dt' \dot{a}(t') \overline{x^2(t')} + \frac{1}{2}\ln[\phi(t) + 2a(0)\psi(t)] - \frac{1}{2}\left(\int_0^t dt' a(t') + 1\right) +$$

$$+ \frac{a(0)}{2a(t)(\phi^2(t) + 2a(0)\phi(t)\psi(t))} \tag{B.3}$$

where $\overline{x^2(t)}$ is calculated from the solution to Eq. (2) (for $b = 0$), which is

$$x(t) = x(0)\phi(t) + \phi(t)\int_0^t dt' \phi(t')^{-1} \xi(t') \tag{B.4}$$

Averaging the square of this expression over the distribution of $\xi(t)$ and over the equilibrium distribution of $x(0) = x_0$ values, given by $P_{eq}(x_0) = \sqrt{a(0)/2\pi D}\exp\left(-a(0)x_0^2/2D\right)$, we find that

$$\overline{x^2(t)} = \frac{D}{a(0)}[\phi^2(t) + 2a(0)\phi(t)\psi(t)] \tag{B.5}$$

Using this result in (B.3), it then follows that $\bar{\sigma}(t)$ is given by Eq. (23).

**FIGURE CAPTION**

1. The time dependence of the function $\bar{\sigma}(t)$ as calculated from Eq. (23) using the definition of the function $a(t)$ as given in Eq. (24a) (first and third panels) and Eq. (24b) (second and fourth panels) at $\alpha = 1$, $\mu = 3$, and $\tau = 10$.

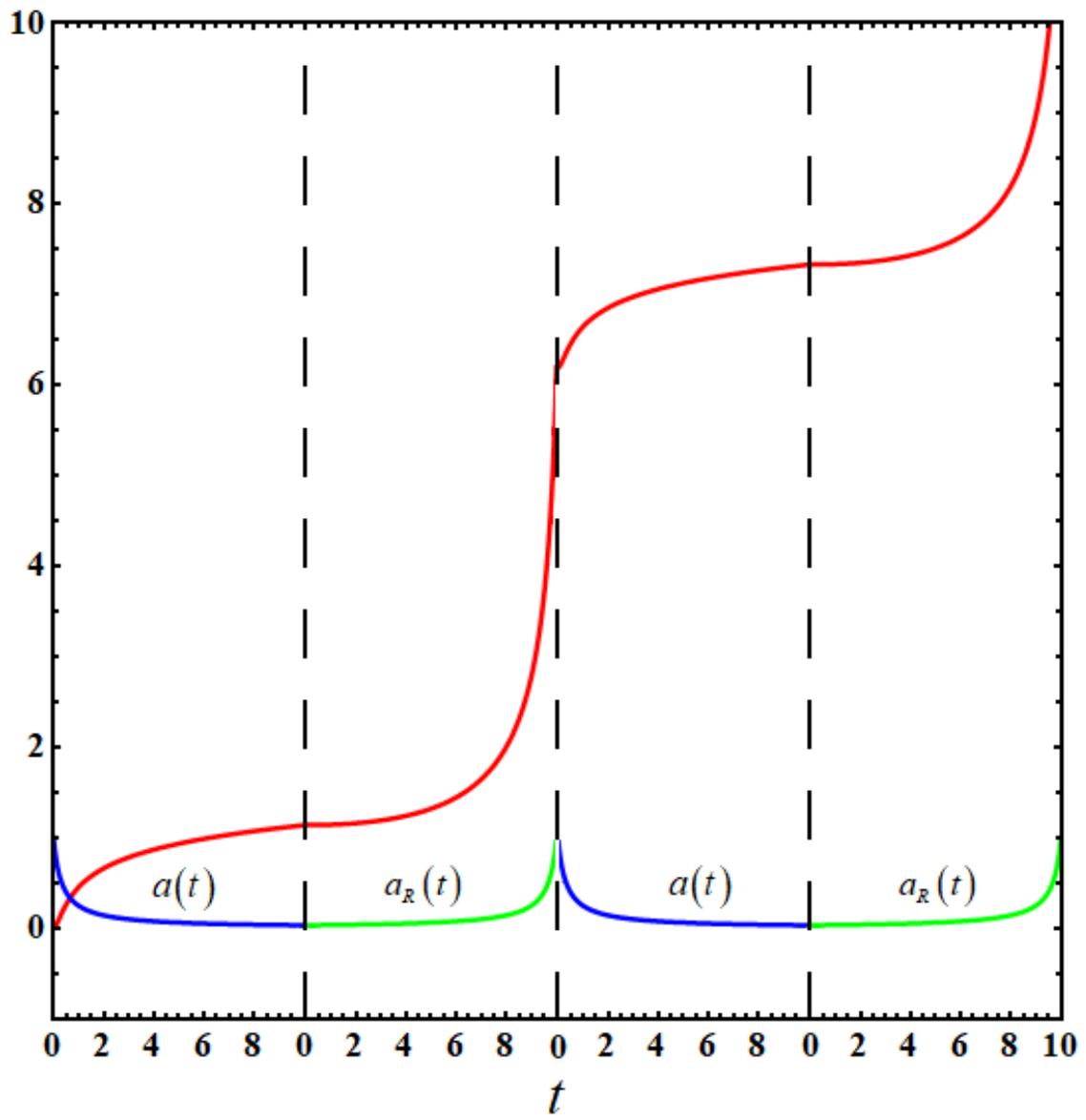

**FIGURE 1**